\documentclass[conference]{IEEEtran}
\IEEEoverridecommandlockouts
\usepackage{amsmath}
\usepackage{graphicx}
\usepackage{multirow}
\usepackage{cite}
\usepackage{booktabs}
\usepackage{subcaption}
\let\svthefootnote\thefootnote
\usepackage[T1]{fontenc}

\def\BibTeX{{\rm B\kern-.05em{\sc i\kern-.025em b}\kern-.08em
    T\kern-.1667em\lower.7ex\hbox{E}\kern-.125emX}}
\begin{document}

\title{Uncertainty-Aware Multi-view Arrhythmia Classification from ECG}

\author{\IEEEauthorblockN{Mohd Ashhad, Sana Rahmani, Mohammed Fayiz, Ali Etemad, Javad Hashemi\\Queen’s University, Canada\\
 \{23qm12; sana.rahmani; 23ntx3; ali.etemad; j.hashemi\}@queensu.ca}}

\maketitle

\begin{abstract}
We propose a deep neural architecture that performs uncertainty-aware multi-view classification of arrhythmia from ECG.
Our method learns two different views (1D and 2D) of single-lead ECG to capture different types of information. We use a fusion technique to reduce the conflict between the different views caused by noise and artifacts in ECG data, thus incorporating uncertainty to obtain stronger final predictions. Our framework contains the following three modules (1) a time-series module to learn the morphological features from ECG; (2) an image-space learning module to learn the spatiotemporal features; and (3) the uncertainty-aware fusion module to fuse the information from the two different views. Experimental results on two real-world datasets demonstrate that our framework not only improves the performance on arrhythmia classification compared to the state-of-the-art but also shows better robustness to noise and artifacts present in ECG.
\end{abstract}
\begin{IEEEkeywords}
ECG, heartbeat classification, uncertainty-aware fusion, multi-view learning
\end{IEEEkeywords}
\let\thefootnote\relax\footnote{Published as a conference paper at IJCNN 2024}
\addtocounter{footnote}{-1}\let\thefootnote\svthefootnote
\section{Introduction}
\label{sec:intro}
Electrocardiography, commonly referred to as ECG, is a crucial medical examination that measures the electrical activity of the heart, playing a crucial role in diagnosing diverse heart conditions and assessing the overall cardiac health of individuals \cite{Attia2019}. In recent years, deep learning has made significant advancements in various fields of medicine, and its role in analyzing and interpreting ECG is no exception \cite{Hannun2019}. The emergence of deep neural networks has enabled the automated analysis of ECG recordings for the identification of diverse abnormalities and arrhythmias \cite{Ahmad2021, Pratiher2022}. Using deep learning in ECG analysis has the potential to significantly affect the detection and treatment of cardiovascular diseases, and as a result, impact the overall efficiency of healthcare systems. 

Like many other signals, ECG contains large amounts of information that can be manifested in various domains or \textit{views}, for instance, raw time-series, spectrograms, and others. As these different views may provide complementary information about ECG signals, it motivates the use of multi-view learning to learn more comprehensive representations from the underlying data \cite{Xu2013}. Upon obtaining representations from different views, an important step is fusing the information to make the final classification decision. This especially becomes more influential in real-world conditions where noise and other artifacts play a role, and different views may have different behaviors in dealing with such conditions \cite{Xu2013}. Thus, the fusion mechanism should ideally consider the confidence levels associated with different views in the process. 

To address this challenge, we present a novel ECG arrhythmia classification framework called Uncertainty-Aware Multi-view Arrhythmia Classification (UAMAC).
Our framework consists of 3 modules. First, a time-series module based on Bidirectional Long Short-Term Memory (BiLSTM) network learns the morphological features of heartbeats. In parallel, an image-space learning module learns the spatiotemporal features from 2D Gramian Angular Field (GAF) representations of ECG heartbeats using a Vision Transformer (ViT). Finally, we use an uncertainty-aware fusion module to aggregate the information learned from the previous two modules effectively while considering the confidence of each module. Comprehensive experiments on two large datasets demonstrate the effectiveness of our framework in ECG heartbeat classification as well as the robustness of our fusion module to noise. 

Our main contributions can be summarized as follows: 
\begin{itemize}

    \item We propose a multi-view deep learning framework for improved ECG arrhythmia detection. Our framework consists of separate modules to learn both the morphological features of heartbeats as well as spatiotemporal features from GAF representations. A fusion module is then used to aggregate the information learned from the previous two modules.

    \item Our model performs uncertainty-aware fusion on the information extracted from different views of ECG and thus better handles discrepancies between the views encountered due to noise and artifacts, achieving state-of-the-art results.

    \item We perform extensive robustness studies to show the effectiveness of our fusion technique as compared to other baselines in the presence of artificial and real-world noise.

\end{itemize}
The paper is structured as follows: Section \ref{sec:Related Work} offers a concise overview of recent advancements in ECG analysis using deep learning. Section \ref{sec:methodology} elaborates our proposed methodology, providing a detailed account of its components. Section \ref{sec:experiments} systematically presents the outcomes of various experiments. 

\begin{figure*}[t]
    \centering
    \includegraphics[width=0.8\textwidth]{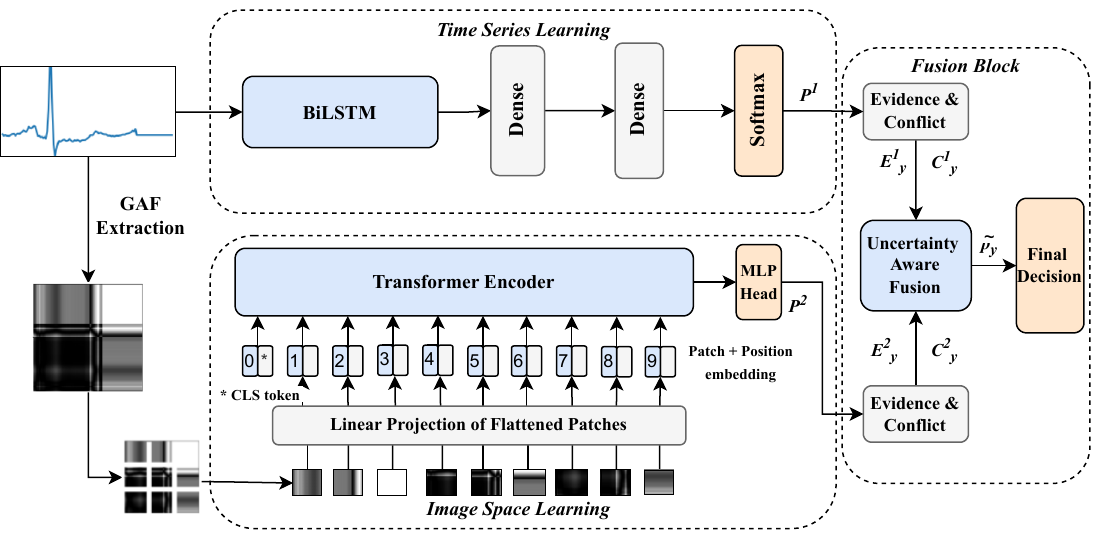}
\caption{Proposed architecture, consisting \textbf{1)} BiLSTM module for time-series learning \textbf{2)} Vision Transformer module for image-space learning on ECG GAF \textbf{3)} The Uncertainty-Aware Fusion Block.}
\label{fig:architecture}
\end{figure*}

\section{Related Work}
\label{sec:Related Work}
Arrhythmia classification from ECG  has been extensively studied through different representations of the ECG signal, such as raw time-series, 2D time-frequency diagrams, and other statistical or morphological features. Using raw ECG data, an S shape reconstruction method was utilized in \cite{Li2022} within 19 layer deep residual network for automatic heartbeat classification. 
In contrast, 2D representations of ECG signals, including spectrograms and scalograms, were incorporated in \cite{KrolJozaga2022} for Atrial Fibrillation (AFib) detection.
In \cite{Hao2019}, spectral-temporal illustrations of ECG were created through wavelet and short-time Fourier transforms to feed Convolutional Neural Networks (CNN) for ECG beat classification. Furthermore, in \cite{Wang2019}, a deep neural network with convolutional and attention layers was developed to extract features at multiple stages for multi-class arrhythmia detection. In \cite{Zvuloni2023}, the performance of feature engineering and deep learning approaches for various classification and regression tasks on ECG data was compared. Results indicate the effectiveness of deep learning for larger datasets while feature engineering approaches performed better on smaller datasets and simpler tasks.

Studies have demonstrated that multi-view ECG learning improves information representation and downstream performance of deep neural networks \cite{Xu2013}. In \cite{Ahmad2021}, GAF, Recurrence Plots, and Markov Transition Field images of ECG were fused, followed by an AlexNet classifier. In \cite{Pratiher2022}, a new variant of the ViT called Dilated Residual ViT (DiResViT) was developed to detect AFib from an ensemble of time-frequency representations of ECG signals. In \cite{Hao2019}, different views of spectral-temporal images were included for beat-to-beat and single-beat information extraction. In \cite{Wang2019}, the use of multi-scale fusion through spatial attention was explored. The proposed model in \cite{Yao2020} integrated a CNN module to learn spatial features, an LSTM module to learn temporal features, and an attention module for multi-class arrhythmia detection from variable length 12-lead ECG data. Furthermore, the framework in \cite{Ge2021} takes advantage of multi-label correlation-guided fusion of features and utilizes multi-scale receptive field fusion modules built with parallel CNN layers for the task of abnormal ECG detection. In \cite{Yang2023}, different leads were considered as distinct views, leveraging the diversity of data captured in a 12-lead ECG for multi-label ECG classification using a multi-scale CNN.

Finally, multiple contrastive methods were used in \cite{Soltanieh2022} and \cite{Soltanieh2023} for self-supervised learning and generalization of multi-lead ECG in the context of In-Distribution and Out-of-Distribution ECG analysis. In the end, we find that although both single-view and multi-view ECG representations have been studied for arrhythmia classification in existing literature, the specific application of ViTs to learn GAF representations still needs to be explored. Furthermore, the use of an uncertainty-aware approach for fusing information from different views with the goal of enhancing robustness to noise has not yet been investigated.

\section{Proposed Method}
\label{sec:methodology}
\subsection{ECG time-series learning}
In order to capture key information from the time domain, we use a BiLSTM network to learn from raw ECG time-series. Given the raw input signal $X$ over $n$ timesteps, the BILSTM layer processes the input sequence $X$ in both forward and backward directions, enabling the network to capture dependencies in both temporal directions effectively. To further enhance the model's capacity to capture complex patterns, we incorporate two additional dense layers with Rectified Linear Unit (ReLU) activation functions. These dense layers contribute to extracting higher-level representations, allowing the model to learn more discriminative features. 
Lastly, a softmax layer is used to obtain class-wise probabilities, denoted as $P^k=\{p^{k}_{j}~|~p^{k}_{j}\in [0,1] , j\in [1,m]\}$ based on view $k$ with $m$ as the class number, which we convert into evidence and conflict,  $E^{k}_{y}$ and $C^{k}_{y}$, respectively formulated as:
\begin{equation}
 E^{k}_{y}=\{p^{k}_{j}~|~p^{k}_{j}\in P^k , j=y\},  
 \label{eq:evidence}
\end{equation}
\begin{equation}
 C^{k}_{y}=\{p^{k}_{j}~|~p^{k}_{j}\in P^k , j\neq y\},
 \label{eq:conflict}
\end{equation}
where $k$ is the view and $y$ is the class index. Accordingly, $E^{k}_{y}$ and $C^{k}_{y}$ will later be used in the fusion stage in order to perform aggregation grounded on the uncertainty of the model with respect to each view (see Fig. \ref{fig:architecture}). 

\subsection{Image-space learning}
We choose GAF as the second view in our method as prior works have shown its effectiveness in capturing effective representations from time-series data \cite{Wang2015, Guan2022, Ahmad2021}. GAF represents the temporal dynamics of a time-series by calculating the pairwise angular differences between its data points and representing these differences as patterns in a 2D image. To obtain the GAF representation, we first apply min-max normalization on the raw ECG $X$ with $n$ time steps, denoted as $\Tilde{X}$, followed by calculating the polar representation. For each data point $\Tilde{x_i}\in\Tilde{X}$, we can calculate the cosine angle (the polar proxy of $\Tilde{x_i}$) as $\phi_i=\arccos \left(\Tilde{x_i}\right)$ and radius (the polar proxy of $i$), as $r_i=\frac{i}{t}$, where constant $t$ regularizes the span of the polar coordinate system for timestep $i$. After mapping $X$ to a polar space, we obtain the GAF representation (Gramian matrix) where each element ${GAF}_{i, j}, \forall i, j \in[1,n]$ is obtained by calculating the cosine of the sum of the angles as ${GAF}_{i, j}=\cos \left(\phi_i+\phi_j\right)$.

Next, our method learns from the GAF representations using ViTs. ViTs have recently emerged as a powerful class of models demonstrating remarkable performance in various computer vision applications \cite{Han2022}. Accordingly, we adopt a ViT-base network \cite{Dosovitskiy2020} for this module of our solution. To account for the single-channel nature of the GAF image representations, we stack $GAF_{i,j}$ three times and resize it to 224$\times$224$\times$3. The resulting matrix is divided into smaller patches of size 16$\times$16$\times$3, which are then passed onto the network. Positional encoding is incorporated to provide spatial information to the model, enabling it to encode the relative patch positions. To incorporate global context for classification, a CLS token is appended to the sequence of patch embeddings, where it acts as an aggregate representation of the entire image to feed the multi-layer perceptron (MLP) head with softmax activation. We convert the output probabilities to evidence and conflict sets as per Eqs. \ref{eq:evidence} and \ref{eq:conflict}.
\begin{table}[t]
    \small
    \centering
    \setlength{\tabcolsep}{2.5pt}
    \caption{Baseline comparison on MIT-BIH and INCART (best performances in bold and second best underlined).}
    {
        \begin{tabular}{lcccc}
        \hline
        \textbf{Method} & \textbf{Dataset} & \textbf{Pre.} & \textbf{Rec.} & \textbf{Acc.(\%)}\\
         \hline
         Kauchee \emph{et al.} \cite{Kachuee2018} &  MIT-BIH & - & -  &93.4\\
         Li \emph{et al.} \cite{Li2023} & MIT-BIH & \underline{95.1} & 93.2 &96.7 \\
         Prakash \emph{et al.} \cite{Prakash2023} & MIT-BIH & 77.3 & 88.5 &96.9  \\
        Shaker \emph{et al.} \cite{Shaker2020} & MIT-BIH & 90.0 & \textbf{97.4} &98.0  \\

        Khan \emph{et al.} \cite{Khan2023} & MIT-BIH & 92.8 & 92.4 &\underline{98.6}  \\
         Ahmad \emph{et al.} \cite{Ahmad2021} & MIT-BIH & 93.0 & 92.0 &\underline{98.6}  \\
         \hline
         \textbf{UAMAC (ours)} & \textbf{MIT-BIH} & \textbf{95.6}& \underline{93.4} &\textbf{98.8} \\
        \bottomrule
        \bottomrule
        He \emph{et al.} \cite{He2018} & INCART & 62.4 & 86.3 & 90.0\\
        Mougoufan \emph{et al.} (CEOP)\cite{Mougoufan2021} & INCART & 75.4 & 77.5 &95.1\\
         Mougoufan \emph{et al.} (PE) \cite{Mougoufan2021} & INCART & 75.4 & 78.2 &95.4\\
         Merdjanovska \emph{et al.}\cite{Merdjanovska2021} & INCART & 65.9 & 75.9 &96.8 \\
        Wu \emph{et al.}\cite{Wu2022} & INCART & \underline{94.2} & \underline{95.0} &\underline{98.1}\\
        \hline
         \textbf{UAMAC (ours)} & \textbf{INCART} & \textbf{96.3}& \textbf{98.0} & \textbf{99.5}\\
                \hline
        \end{tabular}
    }
    \label{tab:baselines}
\end{table}

\subsection{Uncertainty-aware fusion}
Once the two views of the ECG are effectively learned, we propose an uncertainty-aware fusion based on the Dempster-Shafer Theory (DST) \cite{Dempster1968}. This theory is a mathematical framework for reasoning under uncertainty and making decisions based on incomplete or conflicting evidence \cite{Fioretti2001,Jin2022}, which we believe could enhance the outcome of the fusion process given the noise and artifacts often present in ECG datasets \cite{Locher2023}. According to DST, we can combine probability distributions $P^k$ over a set of classes $Z$ obtained from different views if the distributions satisfy the following conditions: (\textit{i}) $P^k(\phi)=0$ where $\phi$ is an empty set, i.e., each sample should belong to at least one class; (\textit{ii}) $\sum_{A \in 2^Z} P^k(A)=1$, where $2^Z$ is the power set of $Z$ (all possible combinations of $Z$). In the case of our problem for ECG-based arrhythmia classification, both conditions hold true given that each sample in our dataset belongs to one and only one class and $P^k$ is a probability distribution generated by a softmax layer and hence adds up to 1. Accordingly, we adopt the Dempster's rule of combination \cite{Fioretti2001} using:
\begin{equation}
    \tilde{p}_y=\frac{E^1_y \cdot E^2_y}{1-C^{1}_{y} \cdot C^{2}_{y} },
\end{equation}
where $E$ and $C$ are the evidence and conflict sets derived according to Eqs. \ref{eq:evidence} and \ref{eq:conflict} based on $P^k$, and $\tilde{p}_y$ is the final uncertainty-aware probability for class $y$. We then obtain the final prediction based on the maximum fused probability.

\section{Experiments and results}
\label{sec:experiments}
\subsection{Implementation details}
All the experiments were performed on two Nvidia 1080 TI GPUs, an Intel i9 9900X processor, and 128 GB of RAM. The models were implemented in Pytorch and were trained for 30 epochs with early stopping and patience of 5, a batch size of 32, and a learning rate of 1e-3 with the Adam optimizer.

\subsection{Datasets}
In this study, we used two real-world ECG datasets, namely MIT-BIH \cite{Moody2001} and INCART \cite{Goldberger2000}. Only the data from lead II was utilized for both datasets. Due to data imbalance, we first created an 80-20 split for training and testing, followed by utilizing the Synthetic Minority Oversampling Technique (SMOTE) \cite{Chawla2002} to balance the class distribution for only the training split. Following, we present a brief description of the datasets.

\subsubsection{MIT-BIH}
The MIT-BIH Arrhythmia Dataset \cite{Moody2001} comprises 48 half-hour segments of two-channel ambulatory ECG recordings that were obtained from 47 patients. From a pool of 4,000 24-hour ambulatory ECG recordings obtained from a varied sample of inpatients and outpatients at Beth Israel Hospital in Boston, 23 recordings were selected at random for inclusion in the dataset. The additional 25 recordings were chosen carefully from the same pool to enrich the dataset with ECG samples having underrepresented arrhythmia. Similar to prior works \cite{Ahmad2021,Khan2023}, we follow the Association for the Advancement of Medical Instrumentation (AAMI) standards to categorize the beats in the dataset into five classes (N, S, V, F, Q).

\subsubsection{INCART Dataset}
Approximately 175,000 annotated heartbeats are available in the St. Petersburg INCART 12-lead Arrhythmia Dataset \cite{Goldberger2000}, which comprises 75 half-hour recordings taken from 32 Holter records of patients undergoing tests for coronary artery disease. Fifteen female and seventeen male subjects provided information for the original records. We followed AAMI standards to categorize the beats from the dataset into 5 classes. However, there were little to no beats available for the F and Q classes. Hence, similar to prior works \cite{Mougoufan2021,Merdjanovska2021}, we use three classes (N, S, V) for training and validation on the INCART dataset.

\subsection{Performance}
In this section, we compare the performance of our model with the state-of-the-art methods on ECG heartbeat classification, as depicted in Table \ref{tab:baselines}. For MIT-BIH, we achieve the highest accuracy of 98.8\% with a precision and recall of 95.6\% and 93.4\%. Compared to \cite{Ahmad2021}, which uses multi-view learning on 2D representations of ECG, we observe improvement in both precision and recall, highlighting the effect of our uncertainty-aware method. Furthermore, by comparing our method, UAMAC, with other baselines \cite{Shaker2020,Li2023,Kachuee2018}, we see an improvement of 0.5\% over the highest competitor \cite{Li2023} in terms of precision. Although our performance on recall is second-best after \cite{Shaker2020}, our overall accuracy and precision are still higher with a margin of 0.2\% and 0.5\%. On the INCART dataset, we achieve an accuracy of 99.5\% with a precision and recall of 96.3\% and 98.0\%, outperforming all the baselines, including the second best method \cite{Wu2022} with an improvement of 1.4\%, 2.1\%, and 3\% on accuracy, precision, and recall.

\begin{table}[t]
   \caption{Performance of different time-series (TS) and image-space (IS) modules.
   } 
    \setlength{\tabcolsep}{5pt}
   \small
   \centering
   \begin{tabular}{lccc|ccc}
   \toprule
    \multirow{2}{*}{\textbf{Methods}} & \multicolumn{3}{c}{\textbf{MIT-BIH}} &\multicolumn{3}{c}{\textbf{INCART}}\\
         & \textbf{Pre.} & \textbf{Rec}. & \textbf{Acc.}  & \textbf{Pre.} & \textbf{Rec.} & \textbf{Acc.}\\
   \midrule
   TS (1D CNN) & 90.0  & 89.0 & 98.1 & 94.2  & 94.0 & 99.0  \\
   TS (Transformer) & 91.0  & 90.2 & 98.4 & 93.6  & 93.8 & 98.9  \\
   \textbf{TS (BiLSTM)} & \textbf{93.4}  & \textbf{92.0} & \textbf{98.6} & \textbf{95.0}  & \textbf{94.6} & \textbf{99.2}  \\
   \midrule
   IS (ResNet18) & 93.0  & \textbf{90.8} & 98.4 & 95.0  & 93.6 & 99.1  \\
   IS (Inception-v3) & 94.0  & 90.0 & \textbf{98.5} & 95.2  & 93.8 & 99.1  \\
   \textbf{IS (ViT base)} & \textbf{94.2}  & 89.7 & \textbf{98.5} & \textbf{95.6}  & \textbf{94.0} & \textbf{99.2}  \\
   \bottomrule
   \end{tabular}
   \label{tab:time ablation}
\end{table}

\begin{table}[t]
   \caption{Ablation experiments.
   } 
    \setlength{\tabcolsep}{5pt}
   \small
   \centering
   \begin{tabular}{lccc|ccc}
   \toprule
    \multirow{2}{*}{\textbf{Methods}} & \multicolumn{3}{c}{\textbf{MIT-BIH}} &\multicolumn{3}{c}{\textbf{INCART}}\\
         & \textbf{Pre.} & \textbf{Rec}. & \textbf{Acc.}  & \textbf{Pre.} & \textbf{Rec.} & \textbf{Acc.}\\
   \midrule
    Time-series only & 93.4  & 92.0 & 98.6 & 95.0  & 94.6 & 99.2  \\
   Image-space only & 94.2  & 89.7 & 98.5 & 95.6  & 94.0 & 99.2\\
   \midrule
   \textbf{UAMAC (ours)} & \textbf{95.6}  & \textbf{93.4} & \textbf{98.8} & \textbf{96.3}  & \textbf{98.0} & \textbf{99.5}  \\
   \bottomrule
   \end{tabular}
   \label{tab:ablation}

\end{table}

\begin{figure}[t]
     \centering
     \begin{subfigure}{0.34\textwidth}
         \centering
         \includegraphics[width=\textwidth]{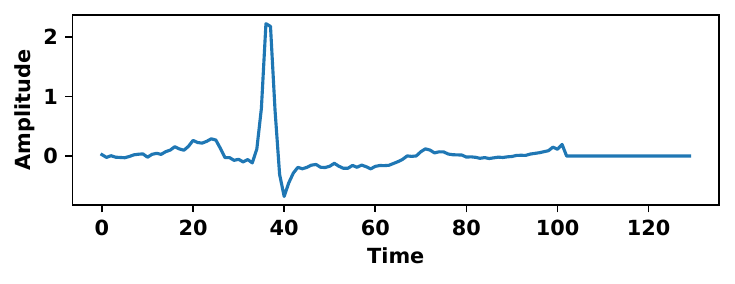}
         \caption{Normal}
         \label{fig:normal}
     \end{subfigure}
     \hfill
     \begin{subfigure}{0.34\textwidth}
         \centering
         \includegraphics[width=\textwidth]{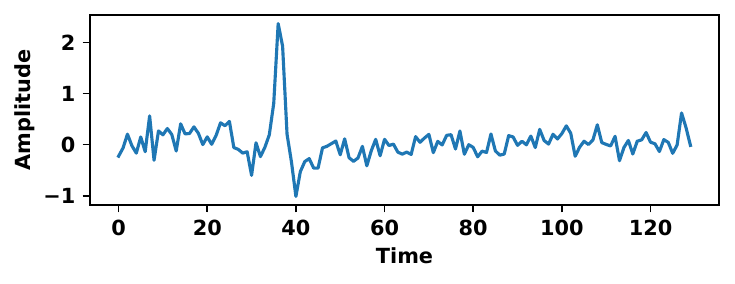}
         \caption{SNR = 5 dB}
         \label{fig:5}
     \end{subfigure}
     \begin{subfigure}{0.34\textwidth}
         \centering
         \includegraphics[width=\textwidth]{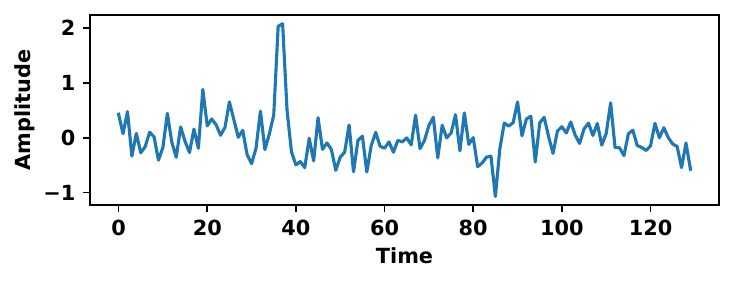}
         \caption{SNR = 0 dB}
         \label{fig:0}
     \end{subfigure}
     \begin{subfigure}{0.34\textwidth}
         \centering
         \includegraphics[width=\textwidth]{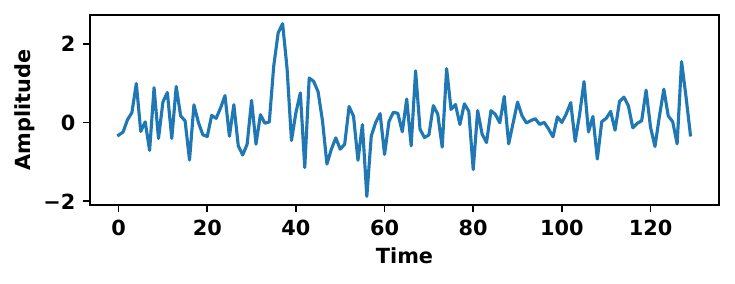}
         \caption{SNR = -5 dB}
         \label{fig:-5}
     \end{subfigure}
        \caption{ECG signals with different levels of noise.}
        \label{fig:noisy ecg}
\end{figure}

\begin{figure*}[t]
     \centering
     \begin{subfigure}{0.25\textwidth}
         \centering
         \includegraphics[width=\textwidth]{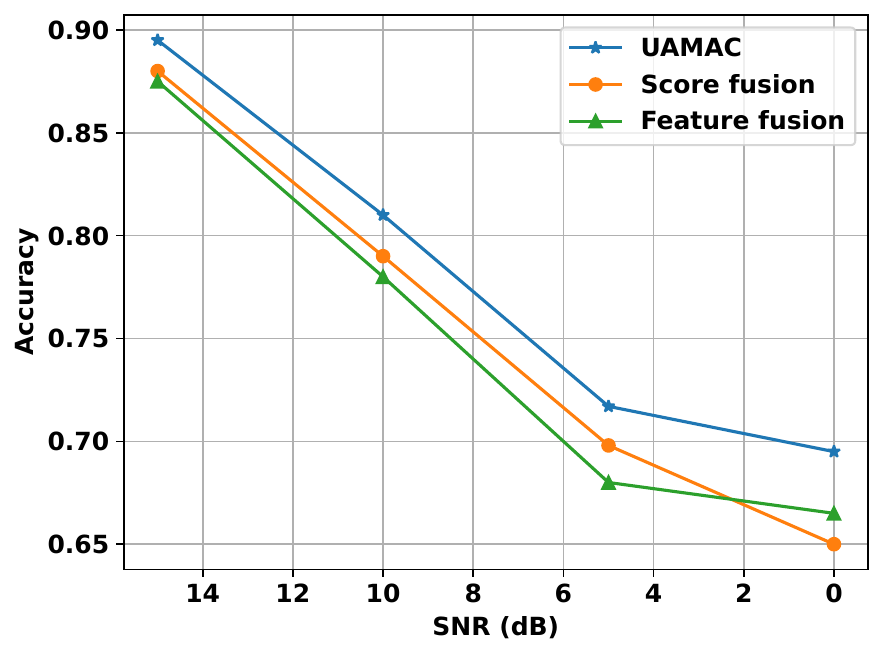}
         \caption{MIT-BIH with BW}
         \vspace{2mm}
     \end{subfigure}
     \begin{subfigure}{0.25\textwidth}
         \centering
         \includegraphics[width=\textwidth]{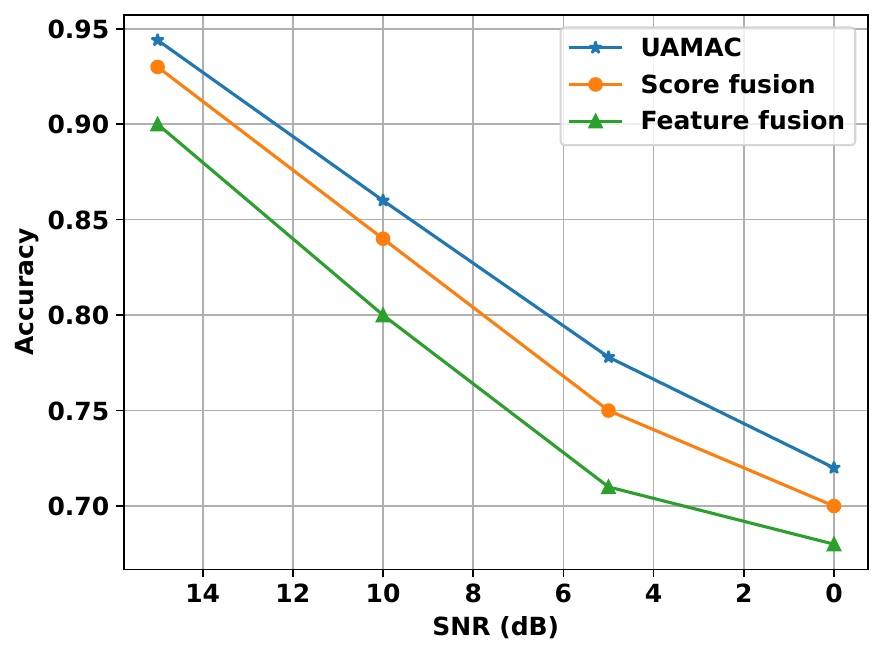}
         \caption{MIT-BIH with EM}
         \vspace{2mm}
     \end{subfigure}
     \begin{subfigure}{0.25\textwidth}
         \centering
         \includegraphics[width=\textwidth]{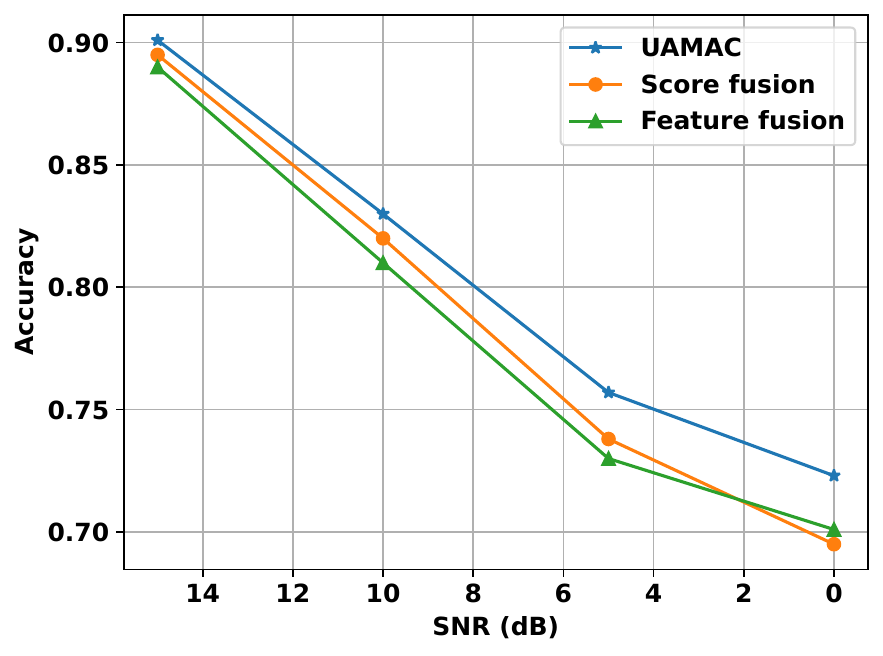}
         \caption{MIT-BIH with MA}
         \vspace{2mm}
     \end{subfigure}
     \begin{subfigure}{0.25\textwidth}
         \centering
         \includegraphics[width=\textwidth]{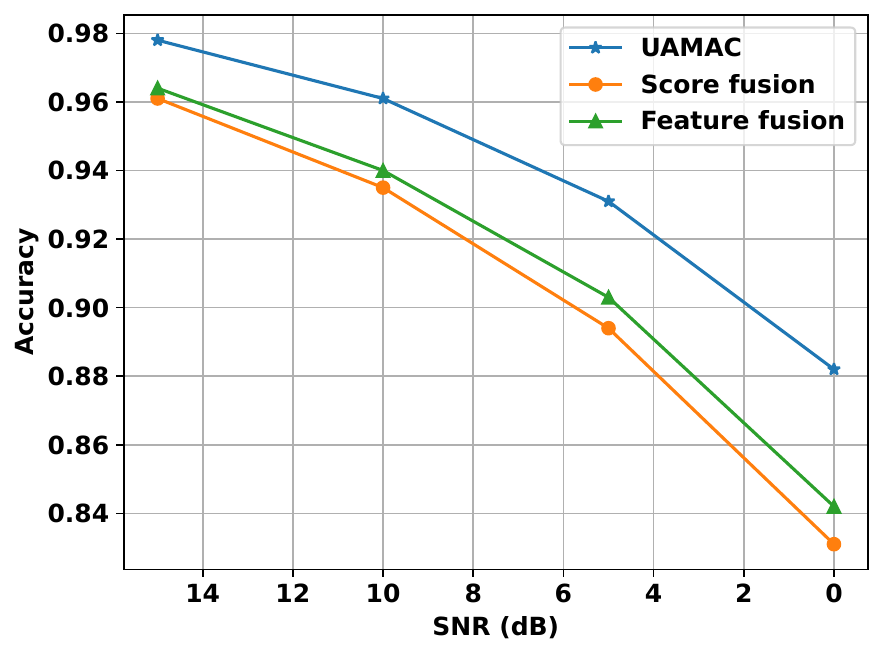}
         \caption{INCART with BW}
         \vspace{2mm}
     \end{subfigure}
     \begin{subfigure}{0.25\textwidth}
         \centering
         \includegraphics[width=\textwidth]{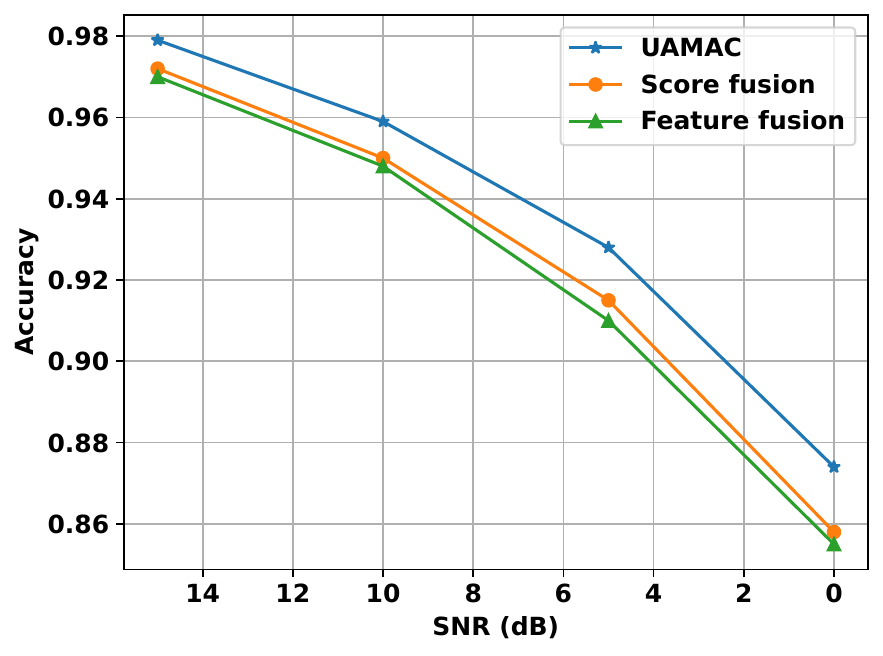}
         \caption{INCART with EM}
         \vspace{2mm}
     \end{subfigure}
     \begin{subfigure}{0.25\textwidth}
         \centering
         \includegraphics[width=\textwidth]{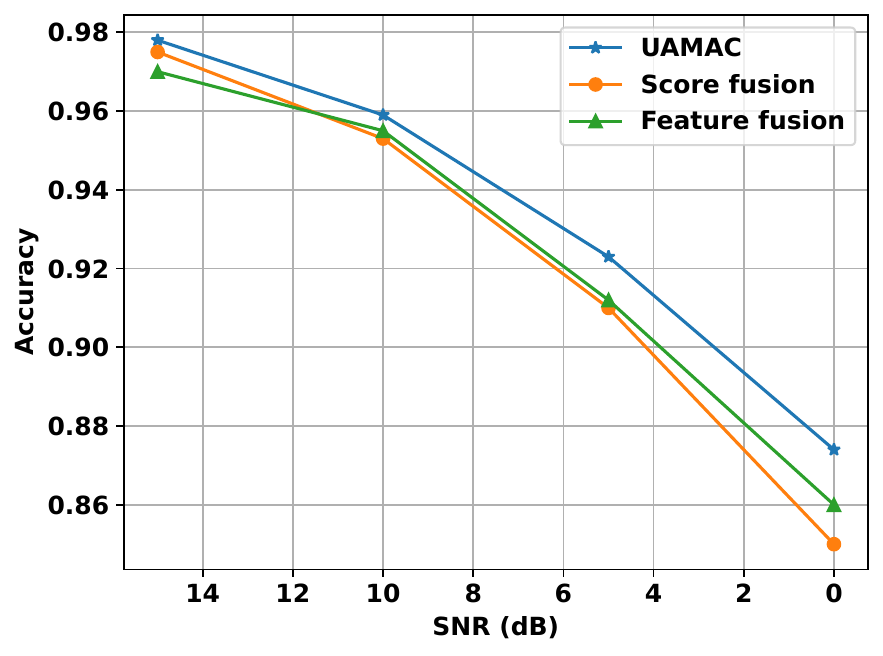}
         \caption{INCART with MA}
         \vspace{2mm}
     \end{subfigure}
     \centering
     \begin{subfigure}{0.25\textwidth}
         \centering
         \includegraphics[width=\textwidth]{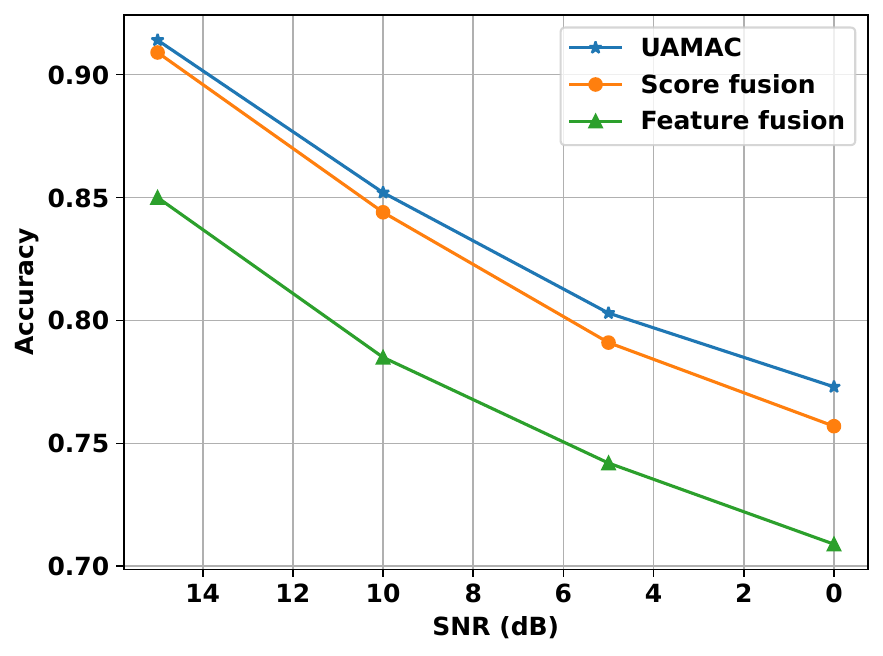}
         \caption{MIT-BIH with AGWN}
         \vspace{2mm}
         \label{fig:mit plot_agwn}
     \end{subfigure}
     \centering
     \begin{subfigure}{0.25\textwidth}
         \centering
         \includegraphics[width=\textwidth]{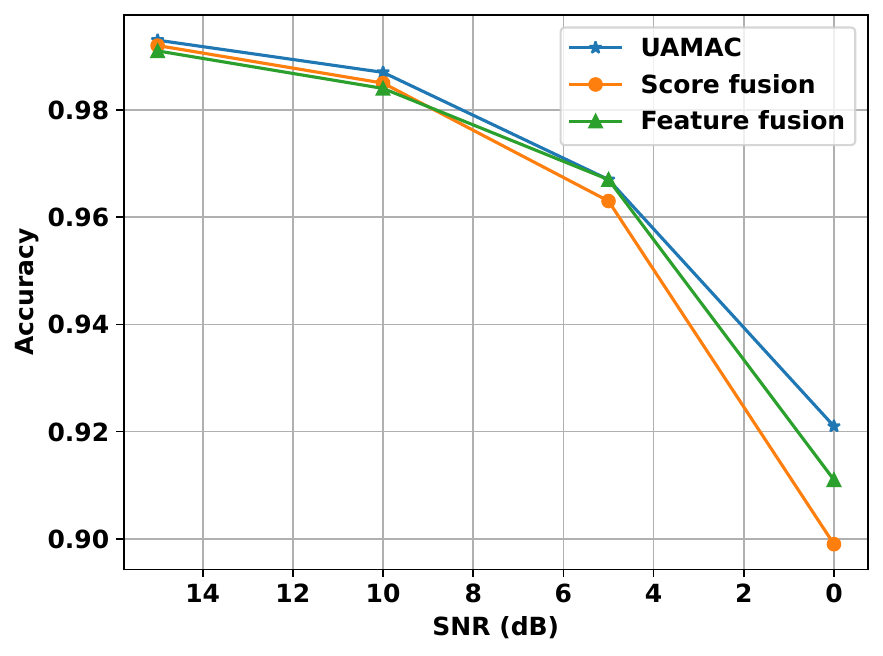}
         \caption{INCART with AGWN}
         \vspace{2mm}
         \label{fig:incart plot_agwn}
     \end{subfigure}
\caption{Performance of our method and different fusion techniques with both artificial and real-world noise from NSTDB.}
\label{fig:noise-real-test}
\end{figure*}

\subsection{Ablation}
To further analyze the performance of our method as well as the contribution of individual modules to the overall performance, we conduct detailed ablation studies. First, we perform the experiments on different views individually (single-view) and present the results in Table \ref{tab:time ablation}. For the time-series module, we investigate different deep architectures, including 1D CNN, Transformer, and BiLSTM, while for the image-space module, we experiment with several popular vision architectures like ResNet18, Inception-v3, and ViT-base. For the 1D CNN, we use three blocks consisting of conv, batch norm, and maxpool layers, while the transformer consists of positional encoding, followed by three multi-head attention blocks with attention, dropout, and normalization layers. For the image-space learning experiments, we use off-the-shelf ResNet-18 and Inception-v3 networks, pre-trained on ImageNet \cite{Deng2009}, where we remove the final layer and fine-tune a classification head using ECG GAFs. As illustrated in Table \ref{tab:time ablation}, the BiLSTM network obtained the best performance for time-series and the ViT for image-space learning, and consequently, they were selected for the final model architecture in UAMAC. Moreover, through the ablation study in Table \ref{tab:ablation}, we observe that eliminating each of the views results in a drop in performance.

\subsection{Robustness to noise}
Finally, we test the impact of the uncertainty-aware fusion module on robustness to noise and artifacts. We compare the performance of UAMAC against the same architecture with score-level and feature-level fusion techniques on the two datasets in the presence of different levels of real-world and artificial noise. First, we added different levels of Additive White Gaussian Noise (AGWN) to the test data and monitored the performance. AGWN is highly controllable and thus very suitable for robustness studies. Fig. \ref{fig:noisy ecg} presents a sample ECG waveform with a few levels of AGWN. Accordingly, we perform the experiments in the range of 15 to 0 dB SNR and avoid the negative dB range as the ECG waveforms are completely convoluted. In Figs. \ref{fig:mit plot_agwn} and \ref{fig:incart plot_agwn}, we observe that our method is more robust to higher levels of AGWN and experiences smaller drops in performance on both datasets. To further test the robustness of our model against real-world ECG noise, we conduct experiments using the MIT-BIH Noise Stress Database (NSTDB) \cite{Moody1984}. The NSTDB comprises three half-hour recordings of noise encountered in ambulatory ECG recordings. The dataset encompasses three distinct types of noise: baseline wander (BW), representing low-frequency variations caused by factors like respiration and body movement; muscle artifacts (MA), caused by high-frequency interference from involuntary muscle contractions; and electrode motion artifacts (EM), signifying disruptions in the contact between skin and electrodes, often induced by patient movement. For our testing, we sampled random excerpts of these noise instances from the NSTDB and added them to the ECG heartbeat data in the test sets of MIT-BIH and INCART and monitored the performance. We introduced varying noise levels to the ECG samples in the range of 15 to 0 dB SNR. As depicted in Fig. \ref{fig:noise-real-test}, our proposed model consistently outperforms other techniques across all three types of noise. This highlights the effectiveness of our approach in real-world scenarios.

\section{Conclusion and Future Work}
In this paper, we propose a new neural architecture called UAMAC for ECG arrhythmia classification. Our model effectively incorporates multiple views from single-lead ECG data through an uncertainty-aware fusion to improve the overall performance. UAMAC contains three modules, including a BiLSTM network for time-series learning, a visual transformer for image-space learning, and finally, an uncertainty-aware fusion to better handle the uncertainty caused by noise and artifacts in ECG data. Extensive experiments on the MIT-BIH and INCART datasets show that our proposed framework outperforms the baselines in the task of ECG arrhythmia classification. Additionally, through detailed experiments with noise, we demonstrate that our fusion approach is more resilient to noise as compared to other commonly used techniques. 
For future work, our study could be extended to take into account other kinds of uncertainty, such as those arising from missing data.

\section*{Acknowledgement}
This work was partially supported by the Natural Sciences and Engineering Research Council of Canada and the Mitacs Globalink Research Internship program.

\bibliographystyle{IEEE}

\end{document}